\documentstyle[twocolumn,aps]{revtex}
\newcommand{\be}{\begin{equation}}
\newcommand{\ee}{\end{equation}}
\newcommand{\ber}{\begin{eqnarray}}
\newcommand{\eer}{\end{eqnarray}}

\begin{document}
\draft

\title{ Laplacian growth as one-dimensional turbulence}
\author{M. B. Hastings and  L. S. Levitov}
\address{Physics Department, Massachusetts Institute of Technology}
\maketitle

\begin{abstract}
A new model of Laplacian stochastic  growth is  formulated
using conformal mappings. The model describes two growth
regimes, stable  and  turbulent,  separated  by  a  sharp  phase
transition.  The first few Fourier components of the mapping
define the web, an envelope of the cluster. The web is  used  to
study the  transition and the dynamics of large-scale features
of the cluster characterized by evolution from macro- to micro-scales.
Also, we derive scaling laws for the cluster size.
   \end{abstract}

The relation of the two kinds of Laplacian dynamics,
stochastic\cite{stochasticLaplacianGrowth} and
deterministic\cite{deterministicLaplacianGrowth}, represents a
very interesting problem. In the stochastic dynamics, such as
diffusion-limited growth\cite{WittenSander}, dielectric
breakdown\cite{Pietronero}, or fracturing, the cluster grows by
particles diffusing and sticking to it, or by emanating
lightning strikes, or cracks, with probabilities determined from
Laplace's equation. The growing cluster is a scale-invariant
fractal object with non-trivial geometric characteristics that
have been extensively explored by a combination of numerical and
analytical methods\cite{Halsey,Pietronero,field-theory,Eckmann}.
In deterministic growth\cite{deterministicLaplacianGrowth},
exemplified by the Hele-Shaw dynamics, the boundary of the
growing object moves with a local velocity determined by
Laplace's equation. In this problem, attention was focused on
the fingering instability, which at small capillary effects
leads to fractal-like patterns\cite{smallCapillaryEffects,link-continuum-limit}
resembling DLA cluster. To draw an explicit relation of the two
large-scale structures, however, turns out to be difficult, even
numerically.

A natural way to proceed would be to study the continuum limit
of stochastic growth, which appears to be non-trivial. Taking
DLA as an example\cite{field-theory}, the arising difficulty is
that the naive continuum limit, understood as taking the
particle size to zero, gives Hele-Shaw dynamics with zero
surface tension\cite{Shraiman}, leading to finite-time
singularities. In that, Laplacian growth has a lot of similarity
with turbulence. The finite-time singularities conjectured for
Euler dynamics of an ideal fluid, are
essential for understanding turbulent flow\cite{EulerSingularity}. The
large-scale spectrum of velocity fluctuations in the flow is
determined by the instability cascade in inertial range, with
viscosity and thermal noise being relevant only on microscale,
where the singularities are resolved. Also, in parallel with
energy conservation in the inertial range of Euler dynamics, it
is found that all integrals of Hele-Shaw dynamics are conserved
in DLA growth as well, up to small fluctuations arising on
microscale\cite{Mineev}. From that, it is natural to conjecture
that the large-scale DLA dynamics is equivalent to Laplacian
contour dynamics, with the finite-time singularities being
resolved on microscale due to noise.

In order to explore this similarity, we propose a new class of
models, continual rather than lattice-like, that provides an
explicit connection between stochastic growth and contour
dynamics. The growth is represented as a random sequence of
conformal maps with memory, which facilitates numerical and
analytical treatment. (It has been shown\cite{Eckmann} that
conformal maps simplify analysis of the mass distribution in the
conventional DLA problem.) In this model there are different
parameter regimes analogous to the low and high Reynolds fluid
dynamics: weakly stochastic, macroscopically stable growth, and
noise-driven turbulent growth. The transition between the two
regimes is sharp. By varying the ``Reynolds'' parameter, one can
model several known growth problems, such as dielectric
breakdown and DLA.

\noindent\underline{\it Stating the problem:}\hskip2mm
One step of Laplacian growth involves attaching a new object to
the cluster with a probability determined by
the solution of Laplace's equation $\nabla^2u=0$, with
$u=0$ at cluster boundary. The probability for the new object to appear
within the interval $dl$ of the boundary is given by $dP=|\nabla
u|dl$. The size of the new object (quantified by its area)
can be \\
  a) constant, as in  DLA;\\
b) proportional to some power of local field, as in dielectric
breakdown. For reasons made clear later we choose to write
the power as $\alpha-2$, so that the
area$\,\sim|\nabla u|^{\alpha-2}$.

Also, there is a freedom to choose different shapes for the
object. To model breakdown and fracturing problems, where an
individual growth step is a lightning strike, or a crack, we
consider growth that involves one-dimensional objects (hereafter
called ``strikes''). For diffusion-controlled growth, it is more
natural to take new objects roughly equal in all dimensions (called
``bumps'').

In the canonical lattice model of dielectric
breakdown\cite{Pietronero} the size of a new object is fixed,
but the sticking probabilities are given by a power law:
$dP=|\nabla u|^{\eta}dl$. To relate it with our model, one
compares local area growth rates, and finds
$\eta=\alpha-1$. Given that, and assuming that universality
classes are determined only by the local area growth law, we
expect that the two models lead to the same macroscopic
properties. Relation to the DLA, known to be a
particular case of dielectric breakdown with
$\eta=1$, is thus expected at $\alpha=2$. However, the relation
of the lattice and the mapping models, although confirmed
by simulations, needs to be explored further.

\noindent\underline{\it  Mapping representation of growth:}\hskip2mm
We use conformal mapping $F(z)$ of the
domain $|z|\ge 1$ to the exterior of the growing cluster.
(Uniqueness up to reparameterization follows from the Riemann
mapping theorem.) To describe the change of the mapping $F$ due
to a single new object added to the cluster we use the mapping
$f_{\lambda,\theta}(z)$ that maps the domain $|z|\ge 1$ onto a
sub-domain by attaching a strike or a bump to the boundary at
$z=e^{i\theta}$ (see insets of Figs. \ref{fig1}, \ref{fig2}).
The strike-mapping $f_{\lambda,\theta}(z)$ is
$e^{i\theta}f_{\lambda}(e^{-i\theta}z)$ where $f_{\lambda}$ is
given by
  \be\label{strike}
\frac{1+\lambda}{2z}(z+1)
\left(z+1+\sqrt{z^2+1-2z \frac{1-\lambda}{1+\lambda}}\right) -1
  \ee
where $\lambda$ is a parameter describing the size of the
strike. Below, we use only mappings with $\lambda\ll 1$, in
which case the strike length is
$2\sqrt\lambda+O(\lambda^{3/2})$. The bump-mapping can be chosen
in several ways. We use
  \be\label{bump}
f_{bump}(z)=z^{1-a}f^a_{strike}(z)\ ,
  \ee
$0<a<1$. The growth is described by composition of mappings:
  \be\label{composition}
F_i(z)= F_{i-1}(f_{\lambda_i,\theta_i} (z))\ ,
  \ee
where the single step mappings $f_{\lambda_i,\theta_i}(z)$ can be
(\ref{strike}) or (\ref{bump}). Note that the order of
the functions in the composition is reversed with respect to the
growth time sequence.

Now we determine the dependence of the parameters $\theta_i$ and
$\lambda_i$ on the growth step. From conformal invariance of the
two-dimensional Laplacian it follows that the random
numbers $\theta_i$ are uniformly distributed in the interval
$0\le\theta\le2\pi$, and uncorrelated. For $\lambda_i$,
the power law relationship of the new object area and local field,
$A\sim|\nabla u|^{\alpha-2}$, gives
  \be\label{lambda_n}
\lambda_i=\lambda_0 |dF_{i-1}/dz|^{-\alpha}_{z=e^{i\theta_i}}\ .
  \ee
(To obtain (\ref{lambda_n}) one just writes the Laplace's problem
solution in terms of the mapping, $u(z)=u_0{\rm Re}\ln F(z)$.)
Using (\ref{lambda_n}) and the composition rule (\ref{composition})
one can grow the object without solving Laplace's equation at
each step.

Using (\ref{composition}) and (\ref{lambda_n}), we simulate the
growth. The examples of
growth patterns obtained for $\alpha=0,\ 1.5,\ 2$ are shown in
Figs.~\ref{fig1},~\ref{fig2},~\ref{fig3}. In the simulation, the
data that represent the structure are the set of angles
$\theta_i$, and the set of $\lambda_i$. For $N$ growth steps,
the algorithm complexity is $N^2$, which compares quite
favorably to the canonical dielectric breakdown
model (e.g., see \cite{breakdowncomplexity}), and comes close to the best
algorithms known for the DLA model.

There is a special parameter value $\alpha=0$ at which the model
(\ref{composition}), (\ref{lambda_n}) is {\it conformally
invariant}, which means that the stochastic dynamics commutes
with arbitrary conformal transformations. As a result, at
$\alpha=0$ the growth does not have any ``memory'': according to
Eq.(\ref{lambda_n}), all $\lambda_i$ are the same, and thus all
mappings $f_{\lambda_i,\theta_i}(z)$ are uncorrelated. As
$\alpha$ increases from $0$ to higher values, the memory effects
become stronger and eventually lead to non-trivial dynamics.

\noindent\underline{\it  The web:}\hskip2mm
To gain some insight, it is useful to consider power series
expansions in $z^{-1}$ for $F_n(z)$ and
$f_{\lambda_n,\theta_n}(z)$:
  \ber\label{pseriesF}
F_n(z)&=&a^{(n)}_1z+ a^{(n)}_0+ a^{(n)}_{-1}z^{-1}+...\\
f_{\lambda_n,\theta_n}(z)&\approx &
(1+\lambda_n)z+\sum\limits_{k=1}^{O(1/\sqrt{\lambda})} 2\lambda_n e^{ik\theta_n}z^{1-k}\ ,
  \label{pseriesf}
  \eer
asymptotically equal to $z+\lambda_n z
(z+e^{i\theta_n})/(z-e^{i\theta_n})$. In the expansion for $f$
we keep only low order terms in $\lambda_n$, since we are
going to work with small $\lambda_n$.  Although to linear order the
cutoff in (\ref{pseriesf}) is infinite, the finite cutoff is
written in explicitly to indicate that $f$ has some smallest scale, beyond which
it is smooth (see below).

By truncating the series (\ref{pseriesF})
one obtains a function that maps the
domain $|z|>1$ to an envelope of the grown cluster (we call it
the {\it web}). The web, even for a relatively small number of
terms, is an accurate representation
of the cluster (see Fig.~\ref{fig3}). The web is
sensitive to the structure of the outer growing part of the
cluster, rather than to the stationary inner region, which
makes the web a useful tool. One observes a change in
the web grown at different $\alpha$. For $0\le\alpha<1$ the
web is rounded, becoming more circular at
large times. On the other hand, at $1<\alpha\le2$ the web is
rough at all times, the roughness scaling as the cluster
size.

An alternative way of imaging the growing cluster used
in the literature is based on the averaged occupancy\cite{link-continuum-limit}.
Its relation to the web represents an interesting problem.

\noindent\underline{\it  The web dynamics:}\hskip2mm
Next, we define {\it web dynamics}. To obtain the web for a
given cluster requires a numerical Fourier transform. Using the
expansions (\ref{pseriesF}), (\ref{pseriesf}) and the rules
(\ref{composition}), (\ref{lambda_n}), we can instead directly
evolve the power series, calculating new coefficents of the
series from old coefficients. To do this,
(\ref{pseriesF}) must be cutoff at some fixed number of terms.
As long as this cutoff is greater than the cutoff in
(\ref{pseriesf}), the web dynamics yields good results. Even
for a relatively small number of terms (say, 20 to 50), it
agrees acceptably with the exact dynamics, both in the
size of the cluster taken as function of time, and in the
overall shape. Several examples of grown webs are shown in the
insets of Fig.~\ref{fig3}. Details on the web dynamics algorithm
will be published elsewhere\cite{tobepublished}.

One advantage of the web dynamics is that it takes much fewer
computer steps than exact dynamics. Also, the web can be used to
study the large-scale limit of the growth dynamics. If for any
reason one is only interested in macroscopic characteristics,
such as the overall growth rate, or the roughness of the cluster
boundary, then the web is a more appropriate instrument than
exact mapping. Finally, the web provides a bridge between exact
dynamics and its continuum limit (see below).

\noindent\underline{\it Taking continuum limit.}\hskip2mm
Consider a continuum limit, in which
the attached objects are infinitesimally small. Then,
the growth is described by {\it deterministic} equation written for
$F(z)$. To derive it we can use the power series
(\ref{pseriesf}). First, substitute the expansion
(\ref{pseriesf}) of $f_{\lambda_n,\theta_n}(z)$ in the recursion
relation (\ref{composition}), and expand in $\lambda_n$:
$F_{n+1}(z)=F_n(z)+\delta F_n(z)$, where
  \be\label{climit1}
\delta F_n(z)= {\partial F_n(z)\over\partial z}\lambda_n z
{z+e^{i\theta_n}\over z-e^{i\theta_n}}\ .
  \ee
Then the continuum limit is implemented by averaging over
$\theta_n$. Taking the integral over $d\theta_n$ and
substituting $\lambda_n=\lambda_0|F_z|^{-\alpha}$, gives
Shraiman-Bensimon equation
  \be\label{climit}
\dot F(z)=\lambda_0 z F_z(z)
\oint |F_z(e^{i\theta_n})|^{-\alpha}
{z+e^{i\theta_n}\over z-e^{i\theta_n}}
{d\theta_n\over2\pi}
  \ee
For $\alpha=2$, which corresponds to constant area of attached object,
i.e., to DLA, we recover the Hele-Shaw
problem which has rich analytic
properties\cite{Shraiman} and can be solved in terms of poles of
$F(z)$ moving within the disk $|z|<1$. It turns out that generic
dynamics leads to singularities occurring at finite time:
$F(z,t)\sim A(t)/(z-w(t))$, $|w(t\to t_0)|\to1$.

Let us compare this behaviour with the web dynamics. The
singularities of the dynamics (\ref{climit}) can be
recognized in the cusp-like features appearing on the web (see
Fig.~\ref{fig3}). After appearing, they become sharper,
but later, instead of developing into real
singularities, they just disappear being
gradually replaced by other similar features nearby.
It is easy to draw a parallel with the  currently  accepted
picture   of  turbulence\cite{EulerSingularity}.  In  the
inertial range, the fluid dynamics gives rise to singular vortex
filaments. In the absence of viscosity these
would develop  into
singularities. However, on  the  viscous  range  the
singularity  is suppressed, and thus a turbulent flow exhibits
a random sequence of  singular-like  structures  replacing  each
other.

In the web, the analog of the spatial scale is the power of $z$.
It can thus be said that sharpening of the cusps on a web
corresponds to the shift to smaller scale. Cusps are suppressed
at microscale set by the largest power of
$z$ in (\ref{pseriesf}). This shortest scale is determined by
the value of $\lambda$, which corresponds to the amount of
noise; in the continuum limit the noise vanishes and the
microscale becomes infinitely small. In the case of web
dynamics, the truncation of (\ref{pseriesF}) may also fix a
microscale. In both cases, for the exact mapping and for the
web, we have singular-like structures developing in the
``inertial range'', where the continuum equation (\ref{climit})
holds, evolving to shorter scales, and disappearing there. It is
thus appropriate to call the long-time dynamics turbulent.

The role of the noise due to the particle discreteness is
unusual. The dynamics without noise would by itself produce
chaos, and even singularities. The noise suppresses
singularities, and makes the system run forever, similar to the
role of viscosity in a turbulent flow. Such behaviour can be
contrasted to other stochastic growth problems, such as the
Kardar-Parisi-Zhang dynamics\cite{KPZ}, or kinetic roughening of
stable Laplacian fronts\cite{Krug}, where all stochastic
properties arise due to the noise. In such problems, naive
continuum limit gives rise to stable dynamics, and thus is
non-problematic.

\noindent\underline{\it Phase transition.}\hskip2mm
As function of $\alpha$, there is a transition at $\alpha=1$
from stable to turbulent growth. The stable growth at $\alpha<1$
is similar for the strike and the bump models: the degree of
roughness of the cluster boundary, scaled in the
cluster size, decreases with time.

The transition at $\alpha=1$ is sharp. To verify this, we plot
the mean square of the fluctuations of the boundary rescaled by
the object size (see Fig.~\ref{transition}). Here, we grew the
object on a periodic strip of finite width
instead of the circular geometry discussed above, and used the
first Fourier component of the mapping as a measure of
fluctuations. As the bump size $\lambda_0$ becomes smaller, the
fluctuations decrease at $\alpha<1$, indicating convergence to
the continuum limit (\ref{climit}), but remain finite at
$\alpha>1$.

These results may not exactly apply to the circular cluster
growth, due to certain differences in DLA growth in the two
geometries. In contrast with the circular growth, in the long
time limit, the growth in the strip geometry is statistically
unchanging, which is why this geometry was selected for the
phase transition study.

The turbulent growth at $\alpha>1$ is sensitive to the model. In
the strike-models the object grows by emanating long chains of
strikes containing almost all the density. On the other hand,
for the bump-models, at $\alpha>1$ the growing object forms a
self-similar fractal cluster, with the fractal dimension varying
with $\alpha$ (see Figs.~\ref{fig2},~\ref{fig3}). The
self-similarity was checked by using gyration radius, which is
found to be a power law function of time over about $10^4$
steps, and by calculating fractal dimension using the box
counting method. The box counting fractal dimension was slightly
lower, but was consistent with the gyration radius power law. (A
similar discrepancy has been noted before\cite{discrepancy}.)
The cluster grown at $\alpha=2$ has properties identical to that
of a DLA cluster. This is expected, since at $\alpha=2$ the new
object size remains roughly constant throughout the growth. So,
the $\alpha=2$ bump models are equivalent to DLA.

\noindent\underline{\it Scaling laws.}\hskip2mm
Given the self-similarity of DLA, it is of interest to study
scaling of the power series coefficients. For that we use the
expansion of $F^{-1}_z=g_0+g_1z^{-1}+...$, the analytic function
whose absolute value gives the local electric field. The mean
square of $g_n$'s are plotted against $n$ on a log-log plot (see
inset to Fig.~\ref{transition}). A renormalization group theory
developed elsewhere\cite{RG} predicts the slope of $-2/5$, also
shown in the plot. The higher Fourier coefficients have a large
effect on the growth rate of the cluster, as discussed below.

The cluster size given by $a^{(n)}_1$ in (\ref{pseriesF}),
by using (\ref{composition}), can be written
in closed form:
  $a^{(n)}_1=\prod (1+\lambda_j)$, \ $0<j\le n$.
For $\alpha<1$, the object is roughly circular, and $\lambda_j$ is determined
by $a_1$.  Noise in the higher components leads to some renormalization
of $\lambda_0\!\to\!\tilde\lambda_0$, and then
$\lambda_j=\tilde\lambda_0 (a_1^{(j-1)})^{-\alpha}$.
This equation has a power law solution:
  \be\label{a_1}
a^{(n)}_1=\left(1+\alpha\tilde\lambda_0n\right)^{1/\alpha}.
  \ee
The same power law exponent can be found from the
circular-symmetric solution to Eq.~(\ref{climit}).

In our simulation, at any $\alpha$ the growth of $a_1$ is indeed
described by a power law, and at $\alpha< 1$, the power law
exponent is equal to $1/\alpha$, as predicted by
Eq.~(\ref{a_1}). However, at $\alpha\ge 1$, the exponent is
somewhat larger than $1/\alpha$, and the deviation increases at larger
$\alpha$. This correction is due to the
growth of the higher Fourier components. Any given one of the
higher Fourier components eventually stabilizes at some average
value, but as the object grows, higher and higher Fourier
components contribute to $\lambda$, and the renormalization of
$\lambda_0$ in Eq.~(\ref{a_1}) increases with time, leading to a
correction in the exponent.

Even at large $\alpha$ it is still true that, on average,
$da_1/dt=a_1 \lambda$, and so, since $a_1$ follows a power law
in time, the average of $\lambda$ scales as $t^{-1}$.
This is equivalent to the electrostatic scaling law
derived by Halsey\cite{Halsey-scaling}

In summary, we studied analytic models of Laplacian stochastic
growth, formulated in terms of conformal mappings, and
characterized two growth regimes: stable ($\alpha<1$) and
turbulent ($\alpha>1$), the transition at $\alpha=1$ being
sharp. By truncating the series expansion of the mapping, we
defined the web, geometric envelope of growing cluster. The web
dynamics is used to demonstrate a relation between transient
features of the turbulent growth and finite-time singularities
of the deterministic problem. The turbulent growth bump-models
are found to be equivalent to the dielectric breakdown and DLA
models.

\acknowledgements
\noindent
This work was initiated by discussions with
Boris Shraiman.

\begin{figure}
\label{fig1}
\noindent
Cluster grown using strike mappings with $\alpha=0$.
{\it Inset:} A strike map applied to three circles of varying radii.
\end{figure}

\begin{figure}
\label{fig2}
\noindent
Cluster grown using bump mappings with $\alpha=1.5$.
{\it Inset:} A bump map applied to three circles of varying radii.
\end{figure}

\begin{figure}
\label{fig3}
\noindent
Cluster grown using bump mappings with $\alpha=2$, and its
web obtained by truncating the cluster mapping series  expansion
at  40  terms;  {\it  Insets 1, 2 :} Examples of growth
using the web dynamics with 20 and 40 terms; {\it Insets 3, 4 :}
For comparison, the webs of the clusters  grown  using  exact
mapping obtained by truncating at 20 and 40 terms.
   \end{figure}

\begin{figure}
\noindent
Boundary roughness measured by the mean square value of the
first Fourier coefficient for three different values of
$\lambda_0$. {\it Inset:}\ Mean squares of Fourier
coefficients of $F^{-1}_z$ scaled by the cluster size $a_1$,
averaged over 50 runs. Theoretical line with the slope $-2/5$ is drawn.
  \label{transition}
\end{figure}


\begin{references}
\bibitem{stochasticLaplacianGrowth}
A.~Erzan, L.~Pietronero, A.~Vespignani,
{\sl Rev. Mod. Phys.} {\bf 67},  545 (1995);
T.~Vitcek, {\sl Fractal Growth Phenomena}, (World Scientific, Singapore, 1992);
H.~E.~Stanley and N.~Ostrowsky, eds.,
{\sl On Growth and Form}, (Nijhoff, Dordrecht, 1986);
\bibitem{deterministicLaplacianGrowth}
P.~G.~Saffman, {\sl J. Fluid Mech.} {\bf 173}, 73 (1986);
D.~Bensimon, et al., {\sl Rev. Mod. Phys.} {\bf 58}, 977 (1986);
\bibitem{WittenSander}
T. A. Witten, and L. M. Sander, {\sl Phys. Rev. Lett.} {\bf 47}, 1400 (1981);
P. Meakin, {\sl Phys. Rev. A} {\bf 27}, 1495 (1983).
\bibitem{Pietronero} L. Niemeyer, L.~Pietronero, and
H.~J.~Wiessmann, Phys. Rev. Lett. {\bf 52}, 1033 (1984);
L.~Pietronero and H.~J.~Wiessmann, J. Stat. Phys. {\bf 36}, 909
(1984).
\bibitem{Halsey}
T.~C.~Halsey and M.~Leibig, {\sl Phys. Rev. A} {\bf 46},  7793 (1992);
T.~C.~Halsey, {\sl Phys. Rev. Lett.} {\bf 72},  1228 (1994).
\bibitem{smallCapillaryEffects}
P.~Tabeling, G.~Zocchi and A.~Libchaber,
{\sl J. Fluid Mech.} {\bf 177}, 67 (1987);
\bibitem{link-continuum-limit}
A.~Arneodo, et al., {\sl Phys. Rev. Lett} {\bf 63}, 984 (1989);
\bibitem{field-theory}
G.~Parisi and Y.~C.~Zhang,
{\sl J. Stat. Phys.} {\bf 41},  ??? (1985);
L.~Peliti, {\sl J. Phys. (Paris)} {\bf 46},  1469 (1985);
Y.~Shapir and Y.~C.~Zhang,
{\sl J. Phys. (Paris) Lett.} {\bf 46},  L529 (1986);
\bibitem{Eckmann}
J.~P.~Eckmann et al.,
{\sl Phys. Rev. Lett.} {\bf 65},  52 (1990);
{\sl Phys. Rev.} {\bf A39},  3185 (1989);
\bibitem{Shraiman}
B.~Shraiman and D.~Bensimon, {\sl Phys. Rev. A}{\bf
30}, 2840 (1984); R.~C.~Ball and M.~Blunt, {\sl Phys. Rev. A}{\bf 39},
3591 (1989).
\bibitem{EulerSingularity}
  S.~Douady, Y.~Couder and M.~E.~Brachet,
{\sl Phys. Rev. Lett.} {\bf 67}, p 983 (1991);\\
A.~Pumir and E.~D.~Siggia,
{\sl Phys. Fluids}, {\bf 30}, 1606 (1987);
{\sl Phys. Fluids}, {\bf A2}, 220 (1990);\\
A.~J.~Chorin, {\sl Comm. Mat. Phys.} {\bf 83}, p 517 (1982);
\bibitem{Mineev}
M.~B.~Minneev-Weinstein, R.~Mainieri,
{\sl Phys. Rev. Lett.} {\bf 72}, 8803 (1994);
\bibitem{breakdowncomplexity}
C.~Amitrano,
{\sl Phys. Rev.} {\bf A39},  6618 (1989);
\bibitem{KPZ} M.~Kardar, G.~Parisi, and Y.~C.~Zhang,
{\sl Phys. Rev. Lett.} {\bf 56},  889 (1986);
\bibitem{Krug} J.~Krug, P.~Meakin,
{\sl Phys. Rev. Lett.} {\bf 66},  703 (1991);
\bibitem{Halsey-scaling}
T.~C.~Halsey, {\sl Phys. Rev. Lett.} {\bf 59},  2067 (1987);
\bibitem{tobepublished} M.~B.~Hastings and L.~S.~Levitov, to be published
\bibitem{discrepancy} F.~Argoul, et al.,
{\sl Phys. Rev. Lett.} {\bf 61}, 2558 (1988);
\bibitem{RG} M.~B.~Hastings, Renormalization theory of stochastic growth,
cond/mat preprint July'96
\end{references}
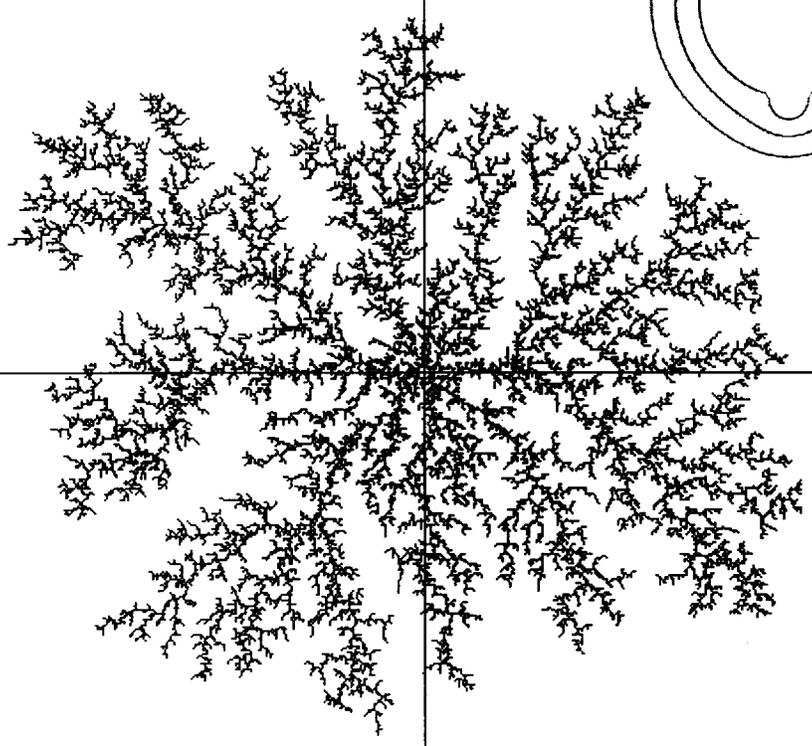
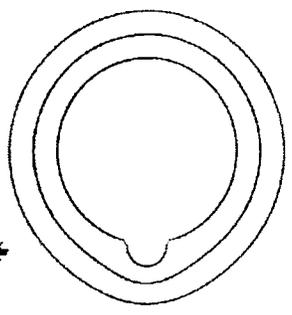
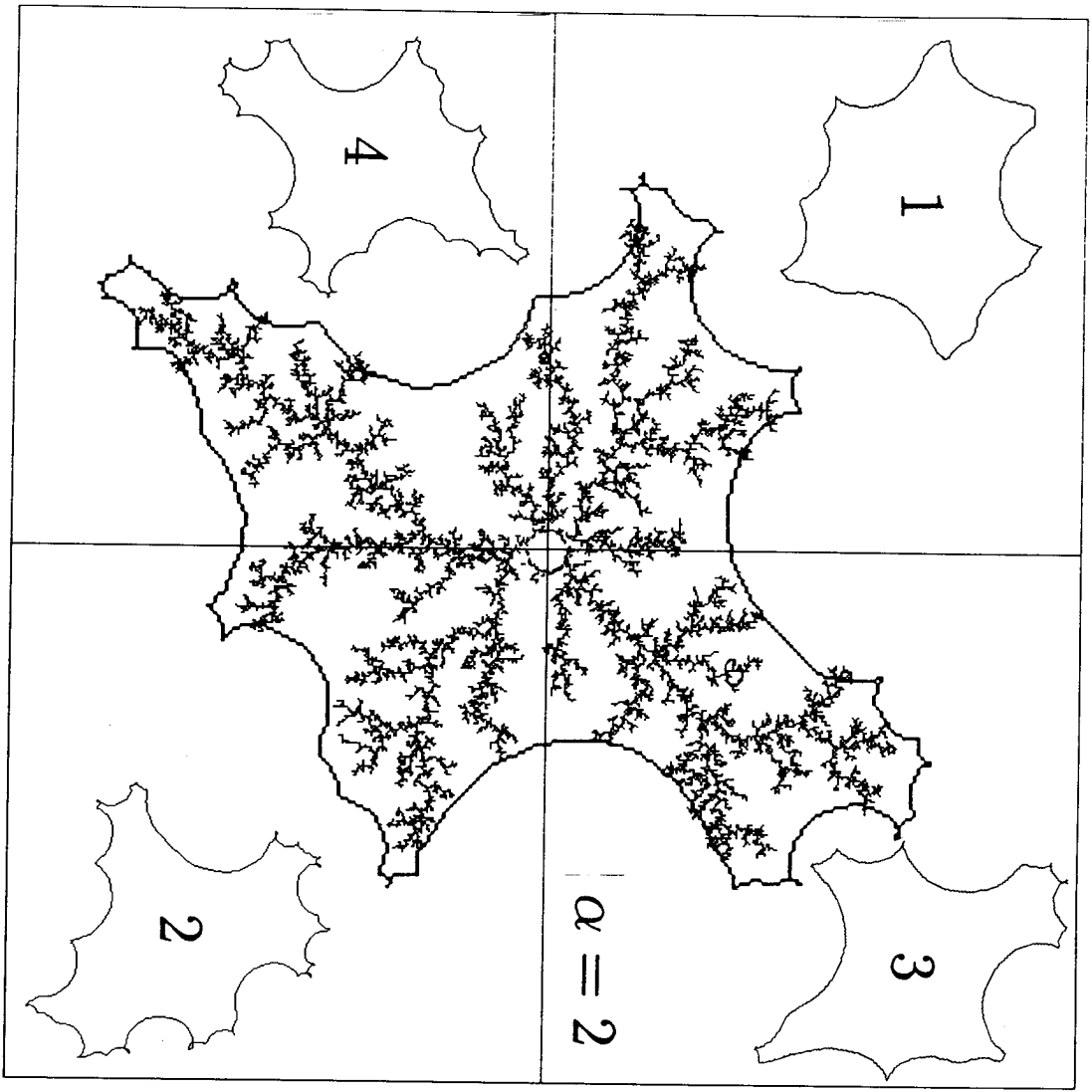
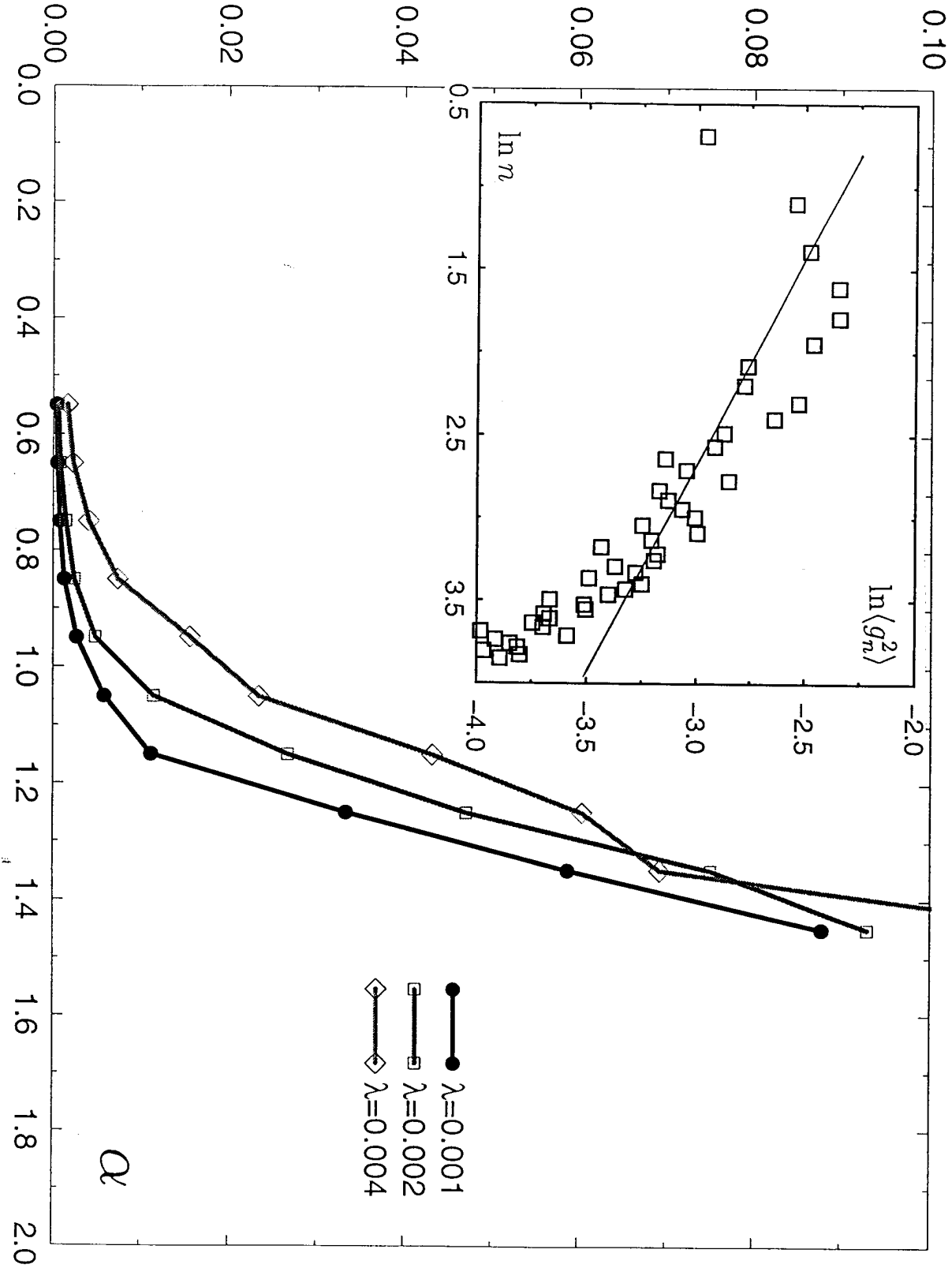
\end{document}